\documentclass[pra,aps,twocolumn,10pt,showpacs,nofootinbib]{revtex4-1}
\usepackage[utf8]{inputenc}
\usepackage[T1]{fontenc}
\usepackage{amsmath}
\usepackage{amsfonts}
\usepackage{amssymb}
\usepackage{mathcmd}
\usepackage{graphicx} 

\usepackage{braket}
\usepackage{bm}
\usepackage{mathrsfs}

\usepackage{tikz}
\usepackage{rotating} 
\usepackage{float}

\newcommand{\mi}{\mathrm{i}}
\newcommand{\me}{\mathrm{e}}
\newcommand{\md}{\mathrm{d}}
\newcommand{\mf}{\mathrm{f}}

\newcommand{\mL}{\mathrm{L}}
\newcommand{\mR}{\mathrm{R}}
\newcommand{\mU}{\mathrm{U}}
\newcommand{\mC}{\mathrm{C}}

\newcommand{\mN}{\mathrm{N}}
\newcommand{\mE}{\mathrm{E}}
\newcommand{\mS}{\mathrm{S}}
\newcommand{\mW}{\mathrm{W}}

\newcommand{\sbm}[1]{\boldsymbol{\mathrm{#1}}}

\begin{document}

\title{Symmetry-protected non-Abelian geometric phases in optical waveguides 
with nonorthogonal modes}
\author{Julien Pinske}
\author{Stefan Scheel}
\email{stefan.scheel@uni-rostock.de}
\affiliation{Institut f\"ur Physik, Universit\"at Rostock, 
Albert-Einstein--Stra{\ss}e 23-24, D-18059 Rostock, Germany}

\date{\today}

\begin{abstract}
The generation of non-Abelian geometric phases from a system of evanescently 
coupled waveguides is extended towards the framework of nonorthogonal 
coupled-mode theory. Here, we study an experimentally feasible tripod 
arrangement of waveguides that contain dark states from which a nontrivial 
$\mathrm{U}(2)$-mixing can be obtained by means of an adiabatic parameter 
variation. We investigate the influence of higher-order contributions beyond 
nearest-neighbour coupling as well as self-coupling on the stability of a 
$\mathrm{U}(3)$-phase generated from an optical tetrapod setup. Our results 
indicate that, despite the mode nonorthogonality, the symmetry of dark states 
protects the geometric evolution of light from distortion.
\end{abstract}
\maketitle

\section{Introduction}
\label{sec:intro}

Abelian as well as non-Abelian gauge theories feature prominently in modern 
theories of fundamental interactions. They also occur frequently in a variety of 
geometric settings and are therefore central to much of modern mathematics. 
Their formulation in terms of gauge fields is deeply connected to the 
notion of a geometric phase. These phase factors arise naturally when 
considering the adiabatic evolution of a state vector in Hilbert 
space \cite{Bohm,Berry,Wilzeck}. 
Besides this deeper insight into geometric and topological notions at 
an experimentally feasible scale, non-Abelian phases (holonomies) are important 
for holonomic and topological quantum computation, where they offer 
parametric robustness to increase stability of the computational 
process \cite{HQC,Pachos}. Consequently, there has been increased 
interest in the study and implementation of artificial gauge fields and symmetry
groups \cite{Cirac}. Successful implementations ranged from single artificial 
gauge fields in photonic \cite{Segev}, superconducting \cite{Abdumal}, and 
atomic systems \cite{Spielman} to experimental simulation of lattice gauge 
theories \cite{Zoller}. 

In recent years, a novel approach was put forward to realise geometric phases 
in terms of integrated photonic structures such as laser-written waveguides in 
fused silica. This have been proven to be a versatile tool box that combines 
the proven capabilities of modern optics with a high degree of interferometric 
stability \cite{Szameit}. For instance, the emergence of a non-Abelian Berry 
phase was observed when injecting coherent states of light into topologically 
guided modes \cite{Chamon,ChamonExp}, whereas other implementations made use 
of a tripod arrangement of evanescently coupled waveguides \cite{Teuber}. In 
Ref.~\cite{Ralph}, a proposal based on a photonic bus mode was studied with the 
view to implement a controlled-NOT gate on single photons. An extension to an 
$N$-pod scheme with arbitrary photon-number states injected was proposed in 
Ref.~\cite{Pinske2020}. However, all the above-mentioned proposals rely on 
power-orthogonality of transverse modes. In general, the deviations from mode 
orthogonality can significantly distort the dynamics of light in coupled-mode 
systems.

In our present work, we overcome the limitations of the weak-coupling regime by 
investigating the properties of geometric phases within the framework of 
nonorthogonal coupled-mode theory (NOCMT). In this setting, the 
nonorthogonality of transverse modes comes to light when decreasing the 
separation between adjacent waveguides. In this regime, the energy 
splitting between an adiabatic subspace and excited states may be large, thus 
favouring the generation of an adiabatic quantum holonomy. We present an 
optical setup for the generation of $\mathrm{U}(2)$-valued and 
$\mathrm{U}(3)$-valued geometric phases arising from tripod and 
tetrapod arrangements of waveguides, respectively, into which coherent light 
is being injected. After an expansion of the system in terms of normal modes, 
an analytical computation of the geometric phase reveals that the symmetry of dark 
states protects the state of light from any distortion. 

The article is structured as follows. In Sec.~\ref{sec:NOCMT}, we review the 
nonorthogonal coupled-mode theory (NOCMT) for a network of weak-index contrast 
waveguides. Sec.~\ref{sec:Tripod} is 
dedicated to the study of a tripod arrangement of waveguides in an 
experimentally feasible setting. We compute its geometric phase 
from a normal mode expansion of the dark states and compare it to a numerical 
propagation of the longitudinal fields, thus quantifying the diabatic error. 
In Sec.~\ref{sec:Tetrapod}, the influence of higher-order coupling and 
self-coupling is illuminated using the example of a $\mathrm{U}(3)$-valued 
geometric phase generated from an optical tetrapod. Finally, 
Sec.~\ref{sec:conclusions} contains a summary as well as some concluding 
remarks. 

\section{Nonorthogonal coupled-mode theory}
\label{sec:NOCMT}

Since the inception of the concept of coupled modes in electromagnetic systems 
\cite{Pierce,Schelkunoff}, coupled-mode theory has become a well-established 
tool in the description of parametric nonlinear devices, waveguide 
structures, optical fibre networks and various other optoelectronic 
structures \cite{Streifer,Chuang,Peall}. 
The starting point of our investigation is the propagation of coherent light 
in a linear, lossless, and isotropic medium patterned with an inhomogeneous 
lattice of $N$ waveguides with refractive-index profile 
$n(\bm{r})=n_{0}+\sum_{j}\Delta n_{j}(\bm{r})$, where $n_{0}$ is the refractive 
index of the host material and $\Delta n_{j}$ is the refractive-index contrast 
of the $j$th waveguide. The dynamics of an electromagnetic wave is determined 
by the inhomogeneous wave equation for the electric field $\bm{E}(\bm{r},t)$
\begin{equation}
 \label{eq:wave}
 \bm{\nabla}^2\bm{E}-\frac{n^2}{c^2}\partial_{t}^2\bm{E}=\bm{E}\cdot \bm{\nabla}\mathrm{ln}\,n^2,
\end{equation}
where $c$ denotes speed of light in vacuum. Under the assumption of a weak 
refractive-index contrast, that is $\bm{\nabla}\mathrm{ln}\,n^2\approx \bm{0}$, 
Eq.~(\ref{eq:wave}) turns into a homogeneous wave equation.
It follows that the Helmholtz equation for the paraxial propagation (slowly 
varying envelope) of an electromagnetic plane wave 
$\bm{E}(\bm{r},t)=\bm{E}(\bm{r})\me^{\mi(\omega t-\beta z)}$ reads
\begin{equation}
 \label{eq:parax}
\left(\bm{\nabla}^2_{\bot}+\lambdabar^{-2}n^2(\bm{r})
-2\mi\beta\partial_{z}-\beta^2\right)\bm{E}(\bm{r})=\bm{0}
\end{equation}
where $\lambdabar=c/\omega$, $z$ denotes the propagation coordinate, $\beta$ is 
the propagation constant, and $\bm{\nabla}^2_{\bot}$ denotes the Laplacian 
with respect to the transverse coordinates $\bm{r}_{\bot}$. In a weak 
index-contrast lattice we have $\Delta n_{j}\ll n_{0}$, hence the vector
character of the Helmholtz equation can be neglected by factoring out a 
constant unit vector $\bm{\nu}$, because polarization effects at the interface 
of a waveguide become unimportant, and scalar wave theory can be 
applied \cite{Gloge}. For the treatment of coupled-mode theory that includes 
high-index contrast waveguides we refer the reader to Ref.~\cite{Bassi}.

In the light of the assumption of weak index contrast, we expand the wave 
packet in terms of tight-binding modes as
\begin{equation}
 \label{eq:Expan}
\bm{E}(\bm{r})=\sum_{j=1}^{N}a_{j}(z)w_{j}(\bm{r})\bm{\nu}, 
\end{equation}
where $w_{j}$ is the transverse field mode that is localised around the 
waveguide $\Delta n_{j}$, and $a_{j}$ is the corresponding longitudinal field 
amplitude determining the dynamics of the light field. The former constitute 
a quasi-complete (complete for divergence free fields) basis in which fields 
with compact support over the transverse plane can be expanded. 
Here we assume that the shape of the transverse field $w_{j}$ remains 
constant throughout the propagation. However, because we consider waveguides 
that vary their position in the transverse plane along the propagation 
direction, the fields $w_{j}$ depend still on $z$. On the other hand, in any 
typical setup the waveguide position changes only very slowly over the 
propagation length (paraxial approximation), so that we can safely assume 
$w_{j}(\bm{r})\approx w_{j}(\bm{r}_{\bot})\me^{\mi\beta_{j}z}$ with an individual 
propagation constant $\beta_{j}$ of the $j$th waveguide. The latter ones can be 
neglected throughout this work by noting that they act merely as offsets when 
considering an identical writing process for each waveguide. 

We assume the transverse fields to satisfy their own Helmholtz equation in 
the $(x,y)$-plane in the absent of the other waveguides, i.e. 
\begin{equation}
\label{eq:eigen}
 \left(\bm{\nabla}_{\bot}^2+\lambdabar^{-2}\left[n_{0}
 +\Delta n_{j}(\bm{r})\right]^2-\beta^2\right)w_{j}(\bm{r}_{\bot})=0.
\end{equation}
This puts the expansion (\ref{eq:Expan}) in the light of a weighted eigenmode 
expansion. We can simplify the above expression by noting that for a weak index 
contrast we have $\left[n_{0}+\Delta n_{j}\right]^2\approx n_{0}^2+2n_{0}\Delta 
n_{j}$ and $\beta^2\approx n_{0}^2/\lambdabar^2$. It thus follows that 
Eq.~(\ref{eq:eigen}) takes the form
\begin{equation}
\label{eq:eigen-2}
 \left(\frac{\lambdabar^2}{2n_{0}}\bm{\nabla}_{\bot}^2+\Delta n_{j}(\bm{r})\right)w_{j}(\bm{r}_{\bot})=0.
\end{equation}
Inserting the ansatz (\ref{eq:Expan}) into Eq.~(\ref{eq:parax}), making use of 
Eq.~(\ref{eq:eigen-2}) and employing the same assumptions as before, we obtain
\begin{equation}
\label{eq:TB}
\sum_{j=1}^{N}\Big(\mi\lambdabar\partial_{z}a_{j}w_{j}
-\sum_{\substack{m=1\\ m\neq j}}^{N}\Delta n_{m}a_{j}w_{j}\Big)=0.
\end{equation}

\begin{figure}[ht]
\centering
\begin{tikzpicture}
\node at (0,0) {\includegraphics[width=0.45\textwidth]{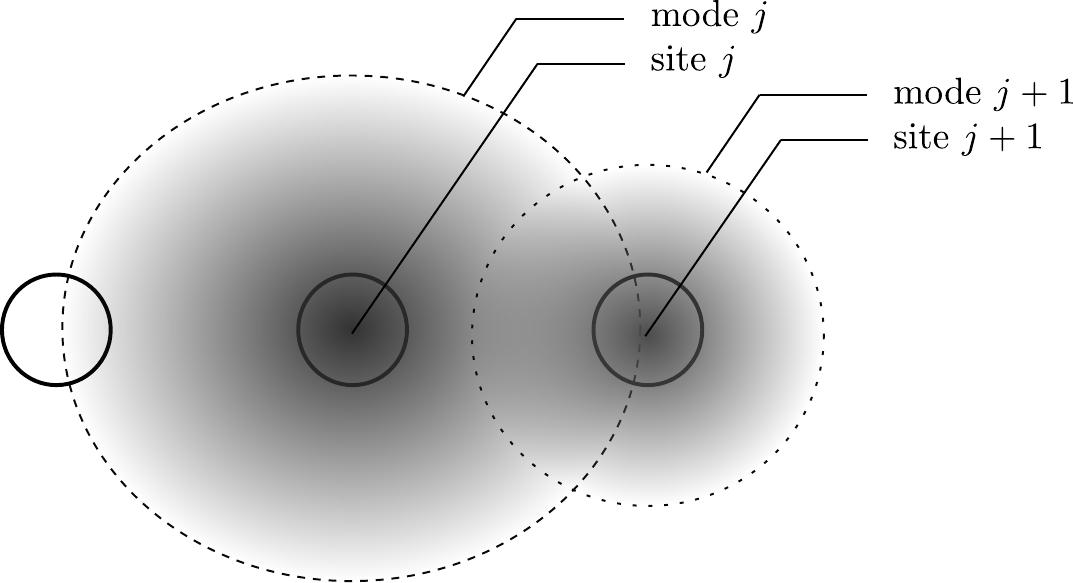}}; 
\node at (-3.6,-1.6) {\includegraphics[width=0.09\textwidth]{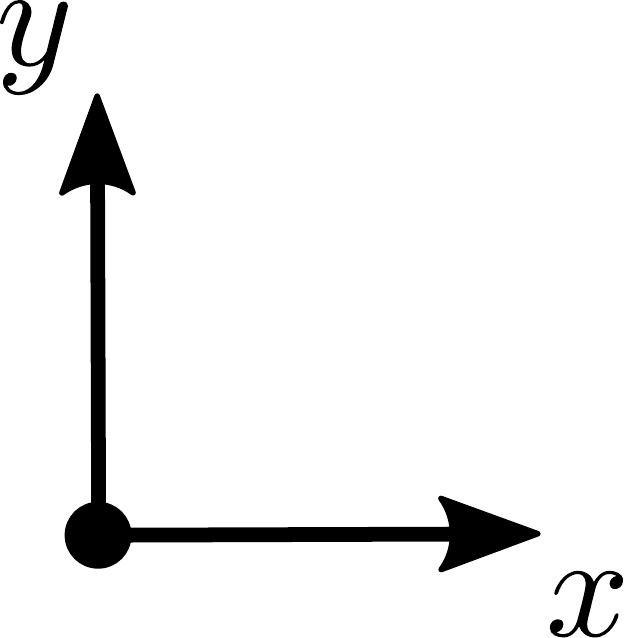}}; 
\end{tikzpicture}
\caption{\label{fig:Modes} Schematic front view of evanescently coupled 
waveguides (black circles). Each site $j$ supports a fundamental transverse mode 
$w_{j}$ (shaded areas). The modes extend towards the nearest neighbouring 
sites.} 
\end{figure}

\begin{figure}[ht]
\centering
\begin{tikzpicture}
\node at (0,0) {\includegraphics[width=0.45\textwidth]{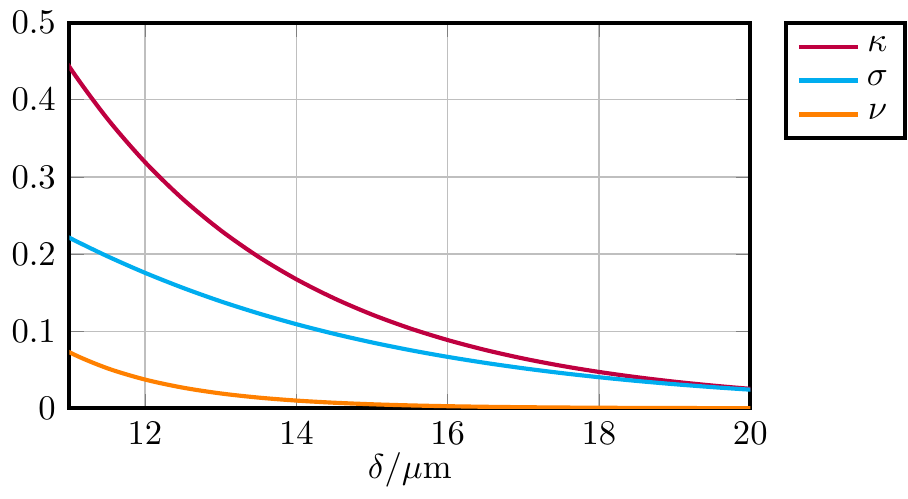}}; 
\end{tikzpicture}
\caption{\label{fig:C&O} Coupling $\kappa$ (in $\mathrm{mm}^{-1}$), overlap 
$\sigma$, and self-coupling $\nu$ (in $\mathrm{mm}^{-1}$) between two adjacent 
waveguides as a function of their separation $\delta$ (measured from the 
center).
The results are shown for cylindrical fibres of radius $R=4.8\,\mu\mathrm{m}$ 
with bulk index $n_{0}=1.452$ and weak contrast $n_{\mathrm{I}}=6.53\cdot 
10^{-4}$ for the inside of each waveguide. The wavelength of the injected light 
beam considered, is $\lambda=633\,\mathrm{nm}$.	
The interpolating functions were obtained by evaluating Eq.~(\ref{eq:C&O}) for 
several values of $\delta$.}
\end{figure}

When contracting Eq.~(\ref{eq:TB}) with $w_{k}^{*}$, we observe that the 
transverse fields overlap with the surrounding waveguides (sites) and their 
modes gives rise to transverse interactions (see Fig.~\ref{fig:Modes}).
Here, we distinguish between two different contributions that occur in 
Eq.~(\ref{eq:TB}) after contraction, namely
\begin{equation}
\label{eq:C&O}
 \begin{split}
 \kappa_{jk}&=\frac{1}{\lambdabar}\sum_{\substack{m=1\\ m \neq j}}^{N} 
 \int_{S_{\infty}}\Delta n_{m} w_{k}^{*}w_{j}\md^2\bm{r}_{\bot},\\
 \sigma_{jk}&=\int_{S_{\infty}} w_{k}^{*}w_{j}\md^2\bm{r}_{\bot},\\
 \end{split}
\end{equation}
where we integrate over the entire $(x,y)$-plane $S_{\infty}$. In 
Eq.~(\ref{eq:C&O}), $\kappa_{jk}\approx 
\frac{1}{\lambdabar}\int_{S_{\infty}}\Delta n_{k} w_{k}^{*}w_{j}\md^2
 \bm{r}_{\bot}$ is the evanescent coupling between waveguides $j$ and $k$, 
while $\sigma_{jk}$ describes the overlap of their respective transverse modes. 
In particular, there is also a self-coupling $\nu_{j}=\kappa_{jj}$ due to 
the presence of other waveguides around the $j$th mode. The latter is 
usually the smallest contribution to the dynamical propagation (\ref{eq:TB}).
This can be seen from Fig.~\ref{fig:C&O} where these quantities were computed 
for two adjacent (identically written) cylindrical waveguides as a function of 
their separation. We considered a scenario in which each waveguide supports only 
its first transverse mode $w_{j}(\bm{r}_{\bot})$, given in terms of Bessel 
functions \cite{FOP}. These assumptions are not too restrictive, as
monomode operation and mode synchronism are often required as operating 
conditions. Furthermore, due to mode overlaps, the transverse modes form a 
nonorthogonal basis for describing the light propagation. The nonorthogonality 
can be neglected if the distance between adjacent waveguides is large. Then one 
works in the regime of OCMT, i.e. $\sigma_{jk}\approx0$ for $j\neq k$ 
\cite{Szameit,Haus,Haus2}.

With the definitions from Eqs.~(\ref{eq:C&O}) at hand, the coupled-mode 
equations become
\begin{equation}
\label{eq:TBE}
\mi \sum_{j=1}^{N}\sigma_{jk}\partial_{z}a_{j}
=\sum_{j=1}^{N}\kappa_{jk}a_{j}.
\end{equation}
This form of the paraxial Helmholtz equation resembles a discrete Schr\"odinger 
equation. Equation~(\ref{eq:TBE}) can be reformulated as a first-order matrix 
differential equation in $z$, with $(\sbm{\Sigma})_{jk}=\sigma_{jk}$ and 
$(\sbm{K})_{jk}=\kappa_{jk}$ being the power and coupling matrix, respectively. 
Because $\sbm{\Sigma}$ and $\sbm{K}$ do not commute in general, the generator 
$\sbm{\Sigma}^{-1}\sbm{K}$ of the dynamics is not necessarily Hermitian, and 
$\sbm{a}^{\dagger}\sbm{a}$ would not be conserved throughout the propagation.
Particularly, we observe that these generators of the NOCMT lie within the 
class of real-valued non-symmetric matrices. Nevertheless, the modified 
intensity distribution $\sbm{a}^{\dagger}\sbm{\Sigma}\sbm{a}$ (POVM 
measurement) remains constant as long as $\sbm{K}$ is Hermitian, that is, 
dissipative effects are negligible as assumed.

The non-Hermitian nature of Eq.~(\ref{eq:TBE}) can even be lifted completely by 
transforming to a set of longitudinal normal modes $\{b_{j}(z)\}_{j=1}^{N}$ such 
that the overall propagation is unitary \cite{Haus}. Even though we did not 
start our derivation from power-orthogonal modes, conservation of energy demands 
that such normal modes always exist. Obviously, in contrast to the waveguide 
mode $a_{j}$, the corresponding normal mode $b_{j}$ will contain contributions 
from adjacent waveguides. To be precise, their relation is given by the 
conserved quantity $\sbm{b}^\dagger\sbm{b}=\sbm{a}^\dagger\sbm{\Sigma}\sbm{a}$.
Moreover, because the power matrix $\sbm{\Sigma}$ is positive definite, there 
exists a (non-unique) matrix $\sbm{Q}$ such that 
$\sbm{\Sigma}=\sbm{Q}^{\dagger}\sbm{Q}$. It follows that the normal modes are 
given by the transformation $\sbm{b}=\sbm{Q}\sbm{a}$. With these preparations, 
Eq.~(\ref{eq:TBE}) can be rewritten as
\begin{equation}
\label{eq:TBH}
 \mi \partial_z \sbm{b}=\sbm{H}\sbm{b},
\end{equation}
where $\sbm{H}$ is given by \cite{Haus}
\begin{equation}
\label{eq:HG}
 \sbm{H}=\left(\sbm{Q}^{-1}\right)^\dagger \sbm{K} \sbm{Q}^{-1},
\end{equation}
which is Hermitian as long as $\sbm{K}=\sbm{K}^{\dagger}$. From similarity 
with Eq.~(\ref{eq:HG}) it follows that $\sbm{\Sigma}^{-1}\sbm{K}$ is always 
diagonalisable with real spectrum. 

\subsection{Adiabatic Propagation}
\label{ssec:NAGP}

While a general state vector $\sbm{\Psi}(z)$ evolves according to 
Eq.~(\ref{eq:TBH}), under the adiabatic assumption any initial preparation 
$\sbm{\Psi}(z_{0})$ in a dark subspace (zero-eigenvalue eigenspace) will be 
mapped to a state 
$\sbm{\Psi}(z_{\mf})=\sbm{U}(z_{0},z_{\mf})\sbm{\Psi}(z_{0})$ 
that is also in the dark subspace. Then, we can expand the state $\sbm{\Psi}$ 
at every instance $z$ in terms of the dark states, i.e. 
$\sbm{\Psi}(z)=\sum_{a}U_{ab}(z)\sbm{d}_{a}(z)$, with initial condition 
$\sbm{\Psi}(z_{0})=\sbm{d}_{a}(z_{0})$. Inserting this ansatz into 
Eq.~(\ref{eq:TBH}) and following Ref.~\cite{Wilzeck}, we obtain
$\left(\sbm{U}^{-1}\partial_{z}\sbm{U}\right)_{ba}=(\sbm{A}_z)_{ab}$
where we used $\sbm{H}\sbm{\Psi}=\sbm{0}$ for states $\sbm{\Psi}$ in the 
dark subspace. Here we defined the adiabatic connection \cite{Chern} as 
$(\sbm{A}_{z})_{ab}=\sbm{d}_{b}^{\dagger}\partial_{z}\sbm{d}_{a}$.                              
A formal solution for $\sbm{U}$ is then given in terms of the matrix exponential
\begin{equation}
\label{eq:holo-Z}
\sbm{U}(z_{0},z_{\mf})=\mathscr{T}\exp\int_{z_{0}}^{z_{\mf}}
\sbm{A}_{z}\,\md z,
\end{equation}
where $\mathscr{T}$ denotes $z$-ordering.

For a collection of coupled waveguides, the $z$-dependence of the system is 
functionally connected to the distances between the waveguides 
$\{\delta_{\mu}(z)\}_{\mu}$ that form local coordinates of an abstract 
parameter space $\mathscr{M}$ (control manifold). Then, the propagation along 
the $z$-direction can equivalently be viewed as a parameter variation along a 
curve $\mathcal{C}:[z_{0},z_{\mf}]\rightarrow\mathscr{M}$. In the following 
sections we are not interested in arbitrary paths, but only those that 
(approximately) form loops in $\mathscr{M}$, i.e. 
$\mathcal{C}(z_{0})=\mathcal{C}(z_{\mf})$.
Population transfer between waveguides associated with such a loop appears to 
be robust against stochastic fluctuations of the control parameter, as the 
transformation $\sbm{U}(z_{0},z_{\mf})$ only depends on the area enclosed by 
the loop $\mathcal{C}$ \cite{Pachos2000}.

\section{Tripod arrangement of waveguides}
\label{sec:Tripod}

\begin{figure}[H]
\centering
\begin{tikzpicture}
\node at (0,0) {\includegraphics[width=0.2\textwidth]{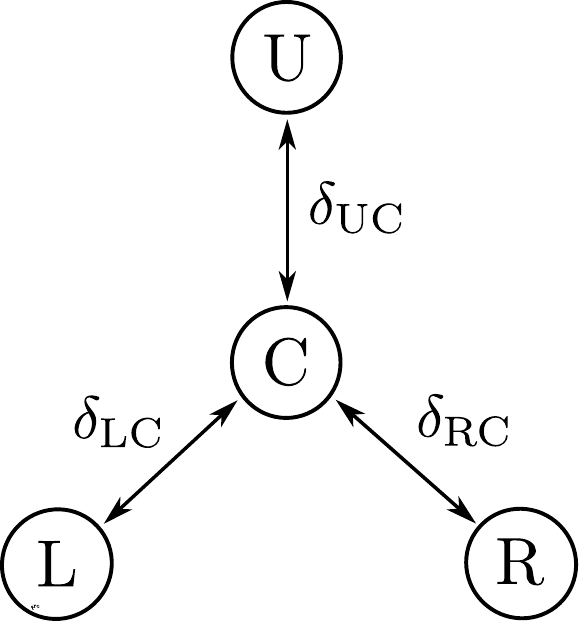}}; 
\end{tikzpicture}
\caption{\label{fig:tripod} Front view of the tripod arrangement, arrows 
indicate interactions between the outer waveguides and the central waveguide. 
Within NOCMT, this includes evanescent coupling as well as the overlap of 
transverse modes. In OCMT, this overlap is neglected. The interaction strength 
depends on the separation $\delta_{i\mC}$ between the corresponding waveguides.}
\end{figure}

In order to examine the properties of geometric phases in relation to 
nonorthogonal modes more closely, we shall focus on a benchmark example
that can be implemented using the current state of technology. A schematic 
representation of the waveguide network under investigation is shown in 
Fig.~\ref{fig:tripod}. There, the outer waveguides $\mL$, $\mR$, and $\mU$ 
interact only implicitly via the central waveguide $\mC$ (nearest-neighbour 
coupling). To be more precise, we consider a situation in which the transverse 
fields of the outer waveguides have couplings and overlaps that can be safely 
disregarded. Then, the coupling matrix of the system is
\begin{equation*}
\label{eq:CS}
 \sbm{K}=
\sum_{j\neq\mC}\kappa_{j\mC}\left(\sbm{w}_{j}^{\dagger}\sbm{w}_{\mC}+\sbm{w}_{
\mC}^{\dagger}\sbm{w}_{j}\right)+\sum_{j}\nu_{j}\sbm{w}_{j}^{\dagger}\sbm{w}_{j}
,
\end{equation*}
where $j$ enumerates the waveguides $\{\mL,\mR,\mU,\mC\}$. Here,
$\{\sbm{w}_{j}\}_{j}$ denotes the standard basis in $\mathbb{C}^4$ that forms 
an orthonormal representation of the transverse eigenmodes of the $j$th 
waveguide. For the moment, we neglect self-coupling on the matrix diagonal, as 
it appears to be the smallest contribution to the overall propagation. 
The corresponding power matrix of the system then reads
\begin{equation}
\label{eq:OS}
\sbm{\Sigma}=
\sum_{j\neq\mC}\sigma_{j\mC}\left(\sbm{w}_{j}^{\dagger}\sbm{w}_{\mC}
+\sbm{w}_{\mC}^{\dagger}\sbm{w}_{j}\right)
+\sum_{j}\sbm{w}_{j}^{\dagger}\sbm{w}_{j},
 \end{equation}
where we assumed normalised transverse modes such that $\sigma_{jj}=1$.
The setup depicted in Fig.~\ref{fig:tripod} is known as a tripod system.
The tripod scheme, arising in different physical settings, is often used as a 
starting point for the generation of $\mathrm{U}(2)$-valued phases, 
see, e.g., Refs.~\cite{Bergmann,Recati,Duan}. 

One possible transformation between waveguide and normal modes can be obtained 
by a Cholesky factorisation of the matrix (\ref{eq:OS}), that reads 
\begin{equation}
\label{eq:Cholesky}
\sbm{Q}=
\sum_{j\neq\mC}\left(\sigma_{j\mC}\sbm{w}_{j}^{\dagger}\sbm{w}_{\mC}+\sbm{w}_{j}
^{\dagger}\sbm{w}_{j}\right)+\sqrt{1-\bm{\sigma}^2}\sbm{w}_{\mC}^{\dagger}\sbm{w
}_{\mC},
\end{equation}
where we introduced the vector 
$\bm{\sigma}=(\sigma_{\mL\mC},\sigma_{\mR\mC},\sigma_{\mU\mC})$.
From the matrix (\ref{eq:Cholesky}) we can compute the Hermitian generator for 
the evolution of normal modes according to Eq.~(\ref{eq:HG}), and we find
\begin{equation*}
\label{eq:HGT3}
\sbm{H}=
\sum_{j\neq\mC}s\kappa_{j\mC}\left(\sbm{w}_{j}^{\dagger}\sbm{w}_{\mC}+\sbm{w}_{
\mC}^{\dagger}\sbm{w}_{j}\right)-2s^2\bm{\sigma}\cdot\bm{\kappa}\,\sbm{w}_{\mC}^
{\dagger}\sbm{w}_{\mC},
\end{equation*}
with $\bm{\kappa}=(\kappa_{\mL\mC},\kappa_{\mR\mC},\kappa_{\mU\mC})$ and 
$s=1/\sqrt{1-\bm{\sigma}^2}$.   
The system's dark states 
\begin{equation}
\label{eq:RDS}
\begin{split}
\sbm{d}_{1}&=\sin\theta\sbm{w}_{\mL}-\cos\theta\sbm{w}_{\mR},\\
\sbm{d}_{2}&=\cos\theta\sin\varphi\sbm{w}_{\mL}
+\sin\theta\sin\varphi\sbm{w}_{\mR}-\cos\varphi\sbm{w}_{\mU},\\
\end{split}
\end{equation}
satisfy $\sbm{H}\sbm{d}_{a}=\sbm{0}$, where we defined
$\theta=\arctan(\kappa_{\mR\mC}/\kappa_{\mL\mC})$ and	
$\varphi=\arctan(\kappa_{\mU\mC}/\sqrt{\kappa_{\mL\mC}^2+\kappa_{
\mR\mC}^2})$. 
Quite remarkably, as long as the second-order (next-nearest-neighbour) 
contributions to the coupling \cite{SzameitSOC} as well as self-coupling are 
negligible, the introduction of mode overlaps does not break the degeneracy of 
the system, and still gives rise to a two-fold degenerate dark subspace similar 
to its counterpart in OCMT \cite{Teuber}. In particular, the dark states of 
$\sbm{H}$ have the same structure as the ones found in OCMT but with their 
coefficients now giving the population of the associated normal mode that 
differ from the individual modes of each waveguide.

With the dark states known, computing the connection $\sbm{A}_{z}$ becomes a 
straightforward task and, thus, the matrix exponential (\ref{eq:holo-Z}) can be 
evaluated analytically to
\begin{equation}
\label{eq:gate}
\sbm{U}(z_{0},z_{\mf})=\begin{pmatrix}
\cos\phi&\sin\phi\\
-\sin\phi&\cos\phi\\
\end{pmatrix},
\end{equation}
written in the basis $\{\sbm{d}_{a}(z)\}_{a}$. Here
\begin{equation}
\label{eq:GP}
 \phi(z_{0},z_{\mf})=\int_{z_{0}}^{z_\mf}\frac{\kappa_{\mU\mC}
\left(\kappa_{\mL\mC}\partial_{z}\kappa_{\mR\mC}-\kappa_{\mR\mC}\partial_{z}
\kappa_{\mL\mC}\right)}
{\sqrt{\kappa_{\mL\mC}^2+\kappa_{\mR\mC}^2+\kappa_{\mU\mC}^2} 
\left(\kappa_{\mL\mC}^2+\kappa_{\mR\mC}^2\right)}\md z
\end{equation}
is a phase factor depending only on the geometry of the associated parameter 
variation $\mathcal{C}$. The holonomy matrix (\ref{eq:gate}) is the same 
unitary transformation as for the case of orthogonal transverse modes 
\cite{Teuber}. This implies that the results of the NOCMT and the OCMT predict 
the same output state as long as the evolution takes place solely in the dark 
subspace. This means that the geometric phase generated from the tripod is 
robust against the nonorthogonal nature of transverse modes. This is a 
remarkable result as, in general, the deviations from mode-orthogonality can 
significantly distort the dynamics of light in a coupled-mode system.   

\subsection{Propagation with Gaussian-shaped geometries}
\label{ssec:AP}

\begin{figure}[H]
\centering
\begin{tikzpicture}
\node at (0,0) {\includegraphics[width=0.45\textwidth]{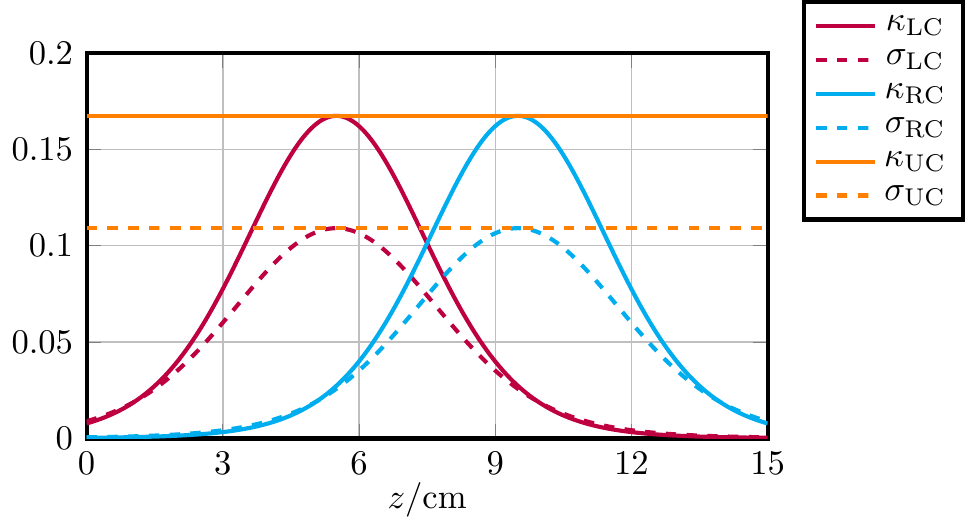}}; 
\end{tikzpicture}
\caption{\label{fig:C&O2} Couplings $\kappa_{i\mC}$ (in $\mathrm{mm}^{-1}$) and 
overlaps $\sigma_{i\mC}$ between the $i$th outer waveguide and the central 
waveguide as a function of the propagation distance $z$ for the case of the 
$z$-dependent waveguide position (\ref{eq:writing}). The chosen parameters
are $\Delta=40\,\mu\mathrm{m}$, $\Delta_{\mU}=14\,\mu\mathrm{m}$, 
$\Omega=26\,\mu\mathrm{m}$, $T=8\,\mathrm{cm}$, and $\tau=2\,\mathrm{cm}$.}
\end{figure}

Here, we present an experimentally feasible parameter loop $\mathcal{C}$ that 
can be realised by current laser-writing techniques \cite{Szameit}. Suppose one
implements the geometry
\begin{equation}
 \label{eq:writing}
 \begin{split}
 \delta_{\mL\mC}(z)&=\Delta-\Omega\,\exp\left(-\frac{\left(z-\overline{z}-\tau\right)^2}{T^2}\right),\\
 \delta_{\mR\mC}(z)&=\Delta-\Omega\,\exp\left(-\frac{\left(z-\overline{z}+\tau\right)^2}{T^2}\right),\\
 \delta_{\mU\mC}(z)&=\Delta_{\mU},\\
 \end{split}
\end{equation}
in which the central waveguide is assumed to remain at the origin of the 
transverse plane, while the separation to the $\mL$ and $\mR$ waveguide changes 
according to a Gaussian function. In Eq.~(\ref{eq:writing}), $\overline{z}$ 
denotes half the total propagation length, $\Delta-\Omega$ is the minimal 
separation between waveguide $\mL$ ($\mR$) and the central waveguide $\mC$, 
with waveguide $\mU$ having constant distance $\Delta_\mU$. Furthermore, $T$ is 
the width parameter, and $\tau$ is the separation of the two Gaussian peaks
from the center at $\overline{z}$.
The relevant coupling strengths and overlaps of the respective waveguides can 
be determined from their mutual separation, shown in Fig.~\ref{fig:C&O}. Given
the variation (\ref{eq:writing}), the respective coupling constants 
and overlaps are shown in Fig.~\ref{fig:C&O2}.

A parameter loop starting at the initial point 
$\bm{\delta}_{0}=(\Delta,\Delta,\Delta_\mU)$ will induce a nontrivial mixing of 
states 
\begin{equation*}
\label{eq:qubits}
\sbm{d}_{1}(\bm{\delta}_{0})\approx\sbm{w}_{\mL},\qquad
\sbm{d}_{2}(\bm{\delta}_{0})\approx\sbm{w}_{\mR}.
\end{equation*}
After integration of the phase factor (\ref{eq:GP}) and subsequent evaluation 
of the transformation (\ref{eq:gate}), this mixing can be given explicitly. 
The adiabatic limit is applicable if the loop $\mathcal{C}$ given by 
Eq.~(\ref{eq:writing}) is traversed slowly enough compared to the energy 
splitting between the dark subspace and the excited bright subspaces. We found 
that the separation between dark and bright states is slightly shifted when 
compared to the eigenvalue splitting known from OCMT, due to the contribution 
from mode overlaps. In order to evaluate to which extent the adiabatic 
approximation is justified in NOCMT, we compute the (expected) gate fidelity 
$0\leq\overline{F}\leq 1$ in terms of input state fidelities
\begin{equation}
 \label{eq:Fid}
 F_{\bm{\Psi}(z_{0})}=|\tilde{\bm{\Psi}}(z_\mf)^{\dagger}\sbm{Q}^{-1}(z_\mf)\sbm{U}(z_\mf)\bm{\Psi}(z_0)|^2,
\end{equation}
with $\sbm{U}(z_\mf)\bm{\Psi}(z_0)$ being the ideal output state predicted by 
the holonomic theory (\ref{eq:gate}) and $\tilde{\bm{\Psi}}(z_\mf)$ obtained 
from a numerical propagation subject to the paraxial Helmholtz equation 
(\ref{eq:TBE}). For the scenario under study, the gate fidelity of the dark 
state ensemble $\{\{1/2,\sbm{w}_\mL\},\{1/2,\sbm{w}_\mR\}\}$ amounts to 
approximately $\overline{F}\approx 98.6\,$\%.

\begin{figure}[H]
\centering
\begin{tikzpicture}
\node at (0,-2.25) {\includegraphics[width=0.35\textwidth]{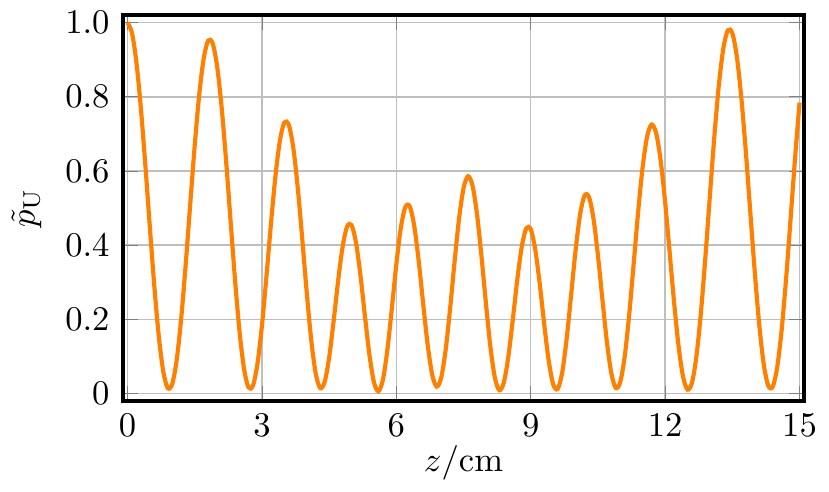}}; 
\node at (0.4,-4.3) {\textbf{(c)} Intensity in site $\mU$ for 
$\sbm{\Psi}(z_{0})=\sbm{w}_{\mU}$.};
\node at (0,-6.75) {\includegraphics[width=0.35\textwidth]{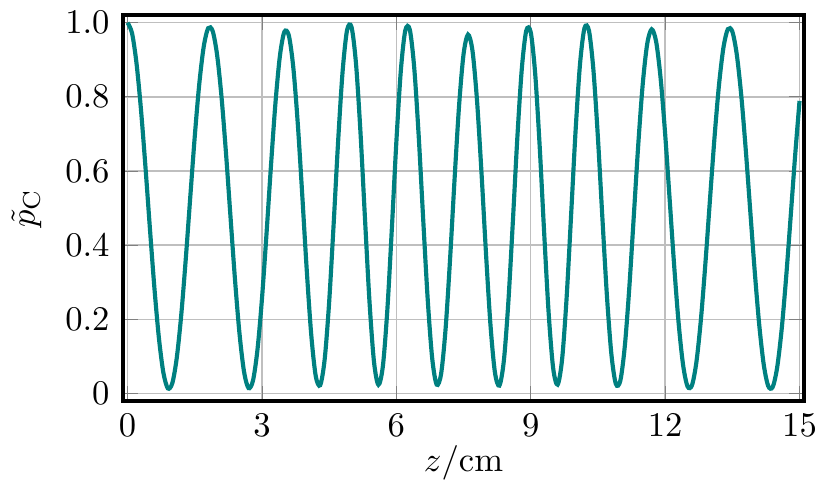}}; 
\node at (0.4,-8.8) {\textbf{(d)} Intensity in site $\mC$ for 
$\sbm{\Psi}(z_{0})=\sbm{w}_{\mC}$.};
\end{tikzpicture}
\caption{\label{fig:Popu} Simulated propagation of $\sbm{\Psi}(z)$ in term of 
waveguide modes [subject to Eq.~(\ref{eq:TBE})] for the waveguide 
position~(\ref{eq:writing}) under investigation. The four plots show the change 
of intensity throughout propagation for different initial input states 
$\{\sbm{w}_{j}\}_{j}$. It can be seen that the bright states ($\mU$ and $\mC$) 
are subject to rapid intensity fluctuations due to deviations from mode orthogonality.
The parameter values are $\Delta=40\,\mu\mathrm{m}$, $\Delta_{\mU}=14\,\mu\mathrm{m}$, 
$\Omega=26\,\mu\mathrm{m}$, $T=8\,\mathrm{cm}$, and $\tau=2\,\mathrm{cm}$.}
\end{figure}

We can conclude that robust adiabatic parallel transport of an initial wave 
packet can be achieved within the range of experimentally feasible setups, 
despite the influence of mode-nonorthogonality. In fact, the expected 
distortion from close-coupling dynamics comes to light when one of the bright 
states $\sbm{w}_{\mU}$ or $\sbm{w}_{\mC}$ is excited. In Fig.~\ref{fig:Popu} we 
compare the population transfer between the waveguides for different initial 
input states. One can clearly see the robustness of the dark states, while the 
excited states are exposed to more rapid population transfer. In fact, when 
adding the bright states to the input ensemble, a gate error emerges that is 
not of diabatic nature, but has its origin in the fact that the parameter 
variation (\ref{eq:writing}) does not form a perfect loop. This matching error 
of input and output basis is rather small for the dark states but gets amplified 
for the bright states due to rapid intensity oscillations, thus leading to a 
substantial decrease in gate fidelity, $\overline{F}\approx 88.9\,$\%.
The relation between mode-nonorthogonality and rapid intensity oscillations was 
experimentally verified in Ref.~\cite{Weimann}. In general, the robustness of 
dark states can be attributed to their form given by Eq.~(\ref{eq:RDS}) which 
is independent of $\sigma_{i\mC}$ and therefore identical to their counterparts 
in OCMT \cite{Teuber}. Hence, despite the overlaps, absence of light in 
waveguide $\mC$ can still be used as a measure of adiabaticity.
\begin{figure}[H]
	\centering
	\begin{tikzpicture}
		\node at (0,6.75) {\includegraphics[width=0.35\textwidth]{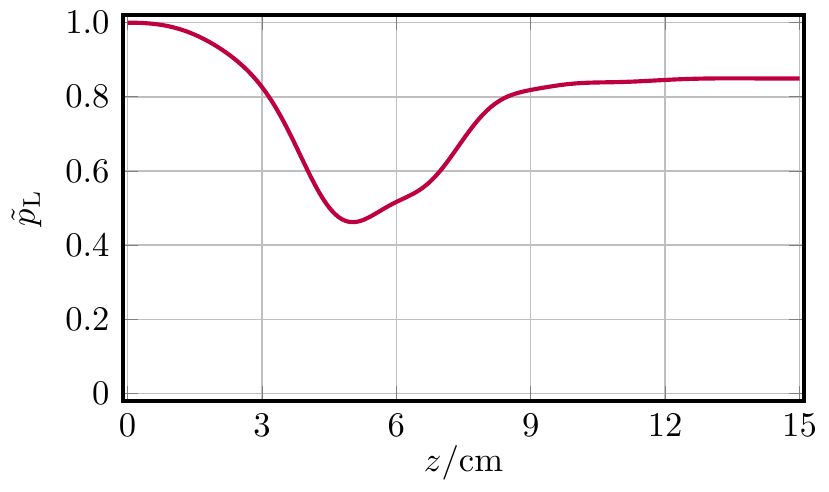}}; 
		\node at (0.4,4.7) {\textbf{(a)} Intensity in site $\mL$ for 
			$\sbm{\Psi}(z_{0})=\sbm{w}_{\mL}$.};
        \node at (0,2.25) {\includegraphics[width=0.35\textwidth]{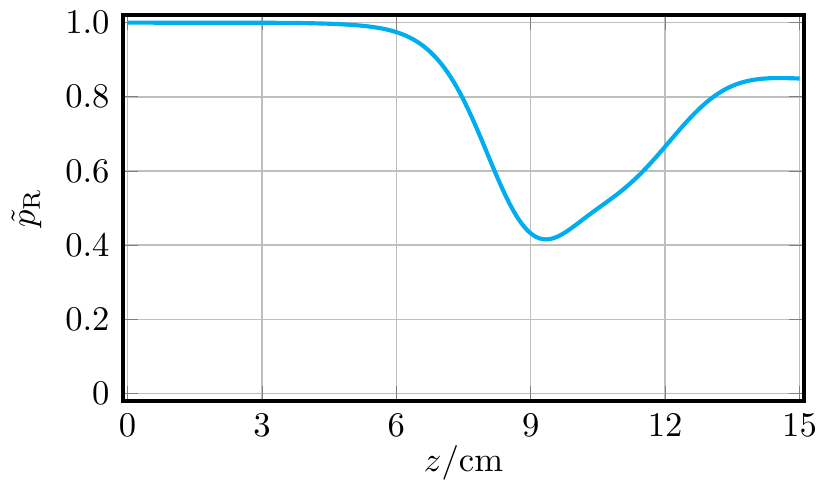}}; 
        \node at (0.4,0.2) {\textbf{(b)} Intensity in site $\mR$ for $\sbm{\Psi}(z_{0})=\sbm{w}_{\mR}$.};
	\end{tikzpicture}
\end{figure}

\section{Higher-order coupling in the tetrapod arrangement}
\label{sec:Tetrapod}

We now turn to the influence of self-coupling and higher-order contributions in 
a waveguide network which had been neglected in our calculations thus far. These
effects potentially break the degeneracy structure of the system under 
investigation. We found numerically that, in the tripod arrangement, these 
contributions are completely negligible for the propagation lengths 
investigated ($|z_\mf-z_0|\leq 15\,\mathrm{cm}$). However, this result depends 
crucially on the distance behaviour shown in Fig.~\ref{fig:C&O}, and does not 
need to be valid beyond step-index waveguides.

\begin{figure}[H]
\centering
\begin{tikzpicture}
\node at (0,0) {\includegraphics[width=0.3\textwidth]{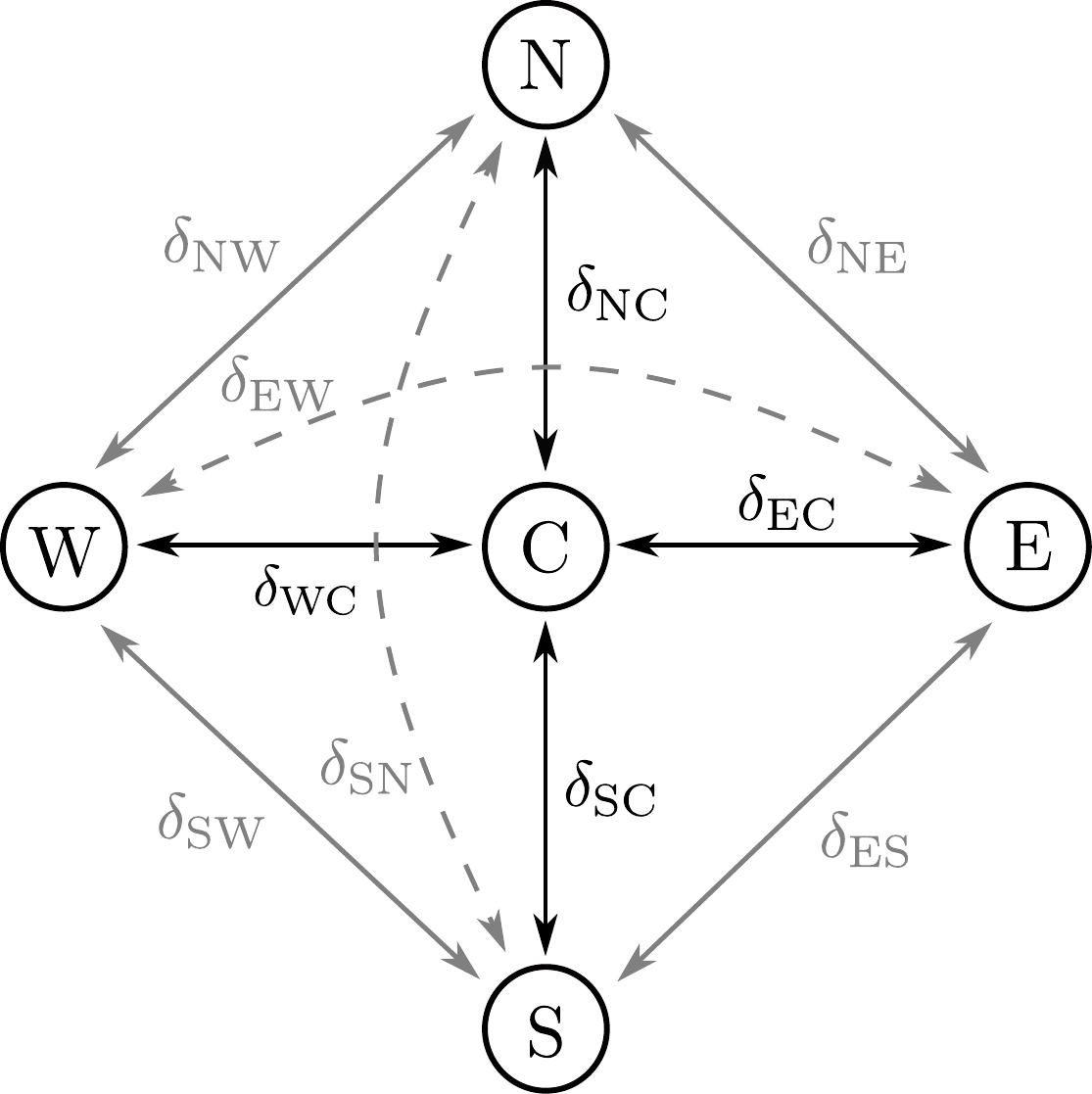}};
\end{tikzpicture}
\caption{\label{fig:tetrapod} Front view of the tetrapod arrangement, arrows 
indicate interaction between the respective waveguides. Within the NOCMT 
this includes evanescent coupling as well as the overlap of transverse modes. 
In the case of OCMT the overlap is neglected. The strength of interaction 
depends on the separation $\delta_{i\mC}$ between the corresponding waveguides.
Black arrows were chosen for nearest-neighbour coupling (first order). Grey 
solid lines are reserved for next-nearest neighbour coupling (second order), 
while grey dotted lines depict coupling between opposite placed waveguides 
(third order).}
\end{figure}

In order to show that coupling beyond nearest-neighbour hopping as 
well as self-coupling can become relevant when the distance between waveguides 
is substantially decreased, we now turn to a tetrapod arrangement depicted in 
Fig.~\ref{fig:tetrapod}. Placing more waveguides around the central site 
will inevitably result in a system having no degeneracy at all. The coupling 
and power matrices of the (ideal) tetrapod system without self-coupling and 
higher-order effects are given by
\begin{equation}
\label{eq:CS4}
\sbm{K}=\sum_{j\neq\mC}\kappa_{j\mC}\left(\sbm{w}_{j}^{\dagger}\sbm{w}_{\mC}
+\sbm{w}_{\mC}^{\dagger}\sbm{w}_{j}\right)
\end{equation}
and
\begin{equation}
\label{eq:OS4}	
\sbm{\Sigma}=
\sum_{j\neq\mC}\sigma_{j\mC}\left(\sbm{w}_{j}^{\dagger}\sbm{w}_{\mC}
+\sbm{w}_{\mC}^{\dagger}\sbm{w}_{j}\right)
+\sum_{j}\sbm{w}_{j}^{\dagger}\sbm{w}_{j},
\end{equation}
where $j$ runs through the alphabet $\{\mN,\mE,\mS,\mW,\mC\}$, and hence 
$\{\sbm{w}_{j}\}_{j}$ might be thought of as the standard basis in 
$\mathbb{C}^{5}$. For this system, the adiabatic propagation of normal modes 
will generate a $\mathrm{U}(3)$-valued mixing within the three-fold degenerate 
dark subspace.

\begin{figure}[h]
\centering
\begin{tikzpicture}
\node at (0,0) {\includegraphics[width=0.45\textwidth]{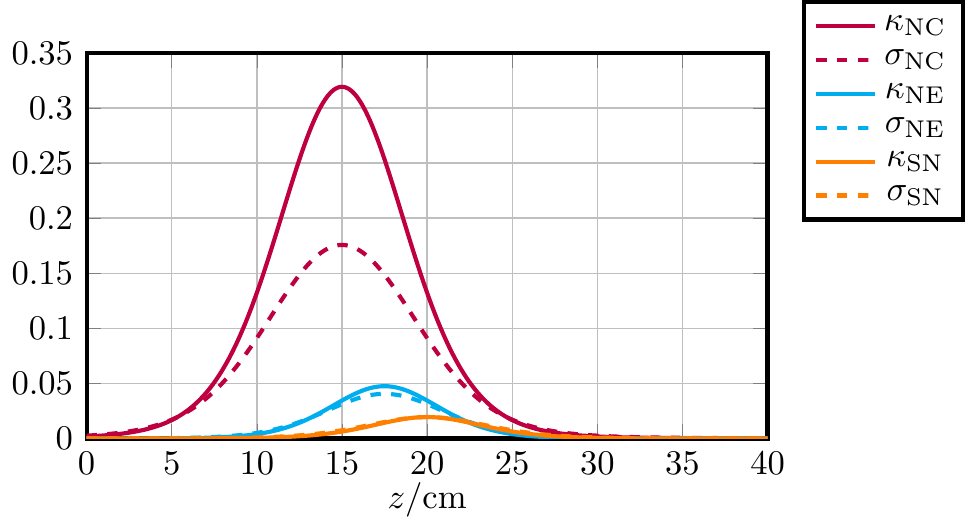}};
\end{tikzpicture}
\caption{\label{fig:tetraCoupling} Couplings (in $\mathrm{mm}^{-1}$) and 
overlaps between the northern ($\mN$) waveguide and $j$th waveguide 
($j\in\{\mC,\mE,\mS\}$) of the tetrapod arrangement as a function of the 
propagation distance $z$ for the waveguide geometry~(\ref{eq:writingTetra}).
Coupling to the central ($\mC$) waveguide is of first order, coupling to the 
eastern ($\mE$) waveguide is of second order, and coupling to the south ($\mS$) 
waveguide is of third order. The relevant parameters are given by 
$\Delta=38\,\mu\mathrm{m}$, $\Delta_{\mW}=12\,\mu\mathrm{m}$, 
$\Omega=26\,\mu\mathrm{m}$, $T=15\,\mathrm{cm}$, and $\tau=5\,\mathrm{cm}$.}
\end{figure}

Higher-order coupling becomes relevant as soon as the propagation length of the 
evolution is substantially increased. In this regime, we can assume 
adiabaticity to apply. Then, the matrices in Eq.~(\ref{eq:CS4}) 
and Eq.~(\ref{eq:OS4}) are no longer sparse, but have only nonvanishing 
entries. The diagonal entries of $\sbm{K}$ now contain the self-couplings 
$\nu_{j}$. In the following, we will study the influence of such contributions 
on a benchmark geometry
\begin{equation}
 \label{eq:writingTetra}
 \begin{split}
 \delta_{\mN\mC}(z)&=\Delta-\Omega\,
 \exp\left(-\frac{\left(z-\overline{z}-\tau\right)^2}{T^2}\right),\\
 \delta_{\mE\mC}(z)&=\Delta-\Omega\,
 \exp\left(-\frac{\left(z-\overline{z}\right)^2}{T^2}\right),\\
 \delta_{\mS\mC}(z)&=\Delta-\Omega\,
 \exp\left(-\frac{\left(z-\overline{z}+\tau\right)^2}{T^2}\right),\\
  \delta_{\mW\mC}(z)&=\Delta_{\mW}.\\
 \end{split}
\end{equation}
In the chosen waveguide geometry, the evolution of coherent light 
approximately starts and ends at the point 
$\bm{\delta}_{0}=(\Delta,\Delta,\Delta,\Delta_{\mW})$, where the 
holonomy will generate a mixing of dark states
\begin{equation*}
\label{eq:qutrits}
\sbm{d}_{1}(\bm{\delta}_{0})\approx\sbm{w}_{\mN},\quad
\sbm{d}_{2}(\bm{\delta}_{0})\approx\sbm{w}_{\mE},\quad
\sbm{d}_{3}(\bm{\delta}_{0})\approx\sbm{w}_{\mS}.
\end{equation*}
Given the geometry (\ref{eq:writingTetra}), we can derive the 
$z$-dependent form of the relevant parameters from Fig.~\ref{fig:C&O}. 
Knowing these parameters allows us, by simple geometric considerations (cf. 
Fig.~\ref{fig:tetrapod}), to construct higher-order couplings and overlaps as 
well. In Fig.~\ref{fig:tetraCoupling}, the coupling strengths of these 
different contributions is shown for the waveguide $\mN$. One observes clearly 
how the higher-order couplings have a substantially smaller influence compared 
to the nearest-neighbour coupling. Nevertheless, for longer propagation lengths 
(e.g. $|z_{\mf}-z_{0}|=40\,\mathrm{cm}$) these contributions will lead to a 
deviation in the intensity distribution from the evolution predicted by the 
propagation through the ideal tetrapod system. 

We also investigated the change of self-coupling $\nu_{j}(z)$ of each 
respective waveguide. There we observed that the order of magnitude of the 
self-coupling in the outer waveguides lies between those of the second-order 
and third-order contributions. However, the self-coupling $\nu_{\mC}$ is 
substantially larger, as the central waveguide mode is closely surrounded by 
the other waveguides.

\begin{figure}[h]
\centering
\begin{tikzpicture}
\node at (0,0) {\includegraphics[width=0.45\textwidth]{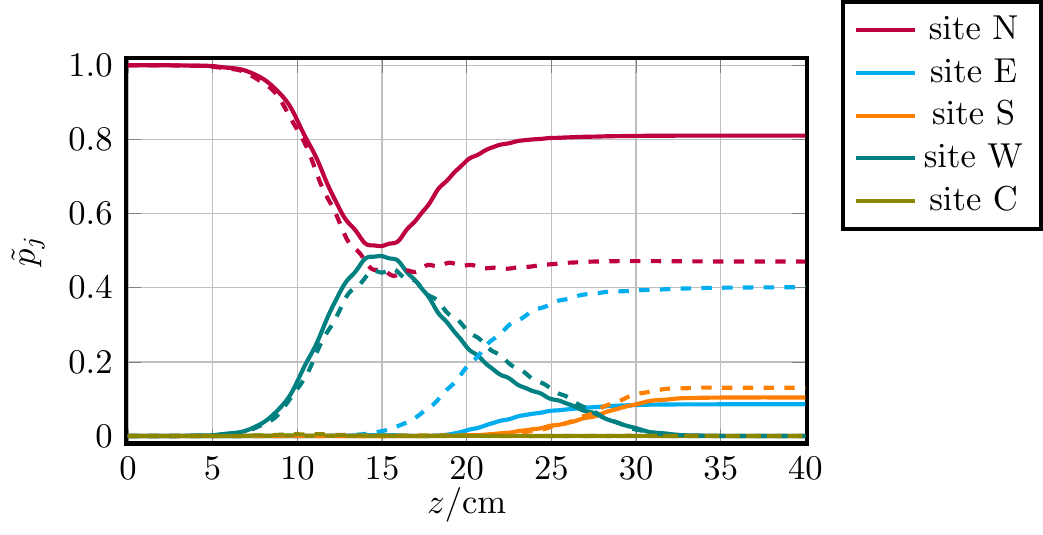}};
\end{tikzpicture}
\caption{\label{fig:tetraIntens} Intensity distribution for the tetrapod 
arrangement subject to the geometry~(\ref{eq:writingTetra}). The solid lines 
belong to the ideal tetrapod [Eq.~(\ref{eq:CS4}) and Eq.~(\ref{eq:OS4})], while 
the dashed lines describe propagation including higher-order couplings as well as self-coupling.
The results were computed via a numerical simulation of Eq.~(\ref{eq:TBE}) for 
an initial wave package $\sbm{w}_{\mN}$, with parameters 
$\Delta=38\,\mu\mathrm{m}$, $\Delta_{\mW}=12\,\mu\mathrm{m}$, 
$\Omega=26\,\mu\mathrm{m}$, $T=15\,\mathrm{cm}$, and $\tau=5\,\mathrm{cm}$ for a 
network of $40\,\mathrm{cm}$ total length.}
\end{figure}

In Fig.~\ref{fig:tetraIntens}, the propagation through the ideal tetrapod 
(solid line) is compared to a propagation including higher-order couplings and 
self-coupling (dashed line). The state was propagated numerically for an 
initial input $\sbm{w}_{\mN}$. Our results show that these additional effects 
(mostly second-order coupling and self-coupling $\nu_{\mC}$) lead to increased 
population transfer between the waveguides. Due to symmetry-breaking, this 
additional population transfer will not be robust against parametric 
fluctuations and mode-nonorthogonalities as it is no longer described by a 
parallel transport in the degenerate dark subspace of the tetrapod.

\section{Conclusion}
\label{sec:conclusions}

In this article, we have shown that the generation of geometric phases in 
integrated photonic waveguide structures can be considered within the framework 
of nonorthogonal coupled-mode theory. General arguments show that, including 
the nonorthogonality of transverse modes leads to a set of tight-binding 
equations governing an evolution in which the conventional intensity 
distribution is not necessarily conserved. This issue was lifted by means of a 
normal mode expansion. An analytical computation of the geometric phase and a 
subsequent evaluation of the intensity distribution for the tripod system showed 
a robustness of the adiabatic parallel transport against deviations from 
mode-orthogonalities as long as adiabaticity holds and higher-order couplings 
are negligible. A subsequent study of the tetrapod arrangement showed that 
higher-order coupling as well as self-coupling can become relevant for longer 
propagation lengths, eventually breaking the degeneracy of the system, thus 
perturbing the generation of a $\mathrm{U}(3)$-valued geometric phase. However, 
for appropriate propagation lengths our numerical simulations of the population 
transfer between the waveguides showed that high fidelity transformations can 
be generated when working in the close-coupling regime.

Our work paves the way for the study of adiabatic parallel transport by 
photonic structures within the close-coupling regime. The symmetry of the dark 
subspace leads to an inherent robustness of geometric phases towards the 
distortion of light by mode-nonorthogonalities. Together with their parametric 
stability, geometric phases can be considered a powerful tool for the generation 
of stable population transfer of light even within the realm of the 
nonorthogonal coupled-mode theory.

\acknowledgments
Financial support by the Deutsche Forschungsgemeinschaft (DFG SCHE 612/6-1) 
is gratefully acknowledged.


\end{document}